\documentclass[]{aa}

\usepackage{xspace}
\usepackage{graphicx}
\usepackage{txfonts}
\usepackage{amsfonts}
\usepackage{amsmath}
\usepackage{amssymb}
\usepackage{subfigure}

\renewcommand{\rho}{\varrho}

\def \gray {$\gamma$-ray\xspace}
\def \grays {$\gamma$-rays\xspace}
\def \flx {photons $\mathrm{cm}^{-2}$ $\mathrm{s}^{-1}$\xspace}

\def \grid {AGILE-\textit{GRID}\xspace}

\def \lat {\textit{Fermi}-LAT\xspace}

\begin{document}

\title{The mid-2016 flaring activity of the flat spectrum radio quasar PKS~2023-07}

\author{G.~Piano\inst{1,2}
\and P.~Munar-Adrover\inst{3}
\and L.~Pacciani\inst{1}
\and P.~Romano\inst{4}
\and S.~Vercellone\inst{4}
\and I.~Donnarumma\inst{5,1}
\and F.~Verrecchia\inst{6,7}
\and L.~Carrasco\inst{8}
\and A.~Porras\inst{8}
\and E.~Recillas\inst{8}
\and M.~Tavani\inst{1,9,10,11}
}
 
\institute{INAF-IAPS, Via del Fosso del Cavaliere 100, I-00133 Roma, Italy
\and CIFS, c/o Dip. Fisica, Univ. di Torino, Via P. Giuria 1, I-10125 Torino, Italy
\and Unitat de F\'{i}sica de les Radiacions, Departament de F\'{i}sica, and CERES-IEEC, Universitat Aut\'{o}noma de Barcelona, E-08193 Bellaterra, Spain
\and INAF, Osservatorio Astronomico di Brera, Via E. Bianchi 46, I-23807 Merate, Italy
\and ASI, Via del Politecnico snc, I-00133 Roma, Italy
\and Space Science Data Center/ASI (SSDC), Via del Politecnico, I-00133 Roma, Italy
\and INAF-OAR, Via Frascati 33, I-00078 Monte Porzio Catone (Roma), Italy
\and Instituto Nacional de Astrofisica Optica y Electronica (INAOE), Apartado Postal 51 y 216, 72000 Puebla, Mexico
\and Astronomia, Accademia Nazionale dei Lincei, Via della Lungara 10, I-00165 Roma, Italy
\and Universit\`{a} ``Tor Vergata'', Dipartimento di Fisica, Via della Ricerca Scientifica 1, I-00133 Roma, Italy
\and Gran Sasso Science Institute, Viale Francesco Crispi 7, I-67100 L’Aquila, Italy
}

\abstract
{Flat spectrum radio quasars (FSRQs) can suffer strong absorption above E = 25/(1+z) GeV, due to gamma-gamma interaction if the emitting region is at sub-parsec scale from the super-massive black hole (SMBH).}
{Gamma-ray flares from these astrophysical sources can be used to investigate the location of the high-energy emission region and the physics of the radiating processes.}
{We present an episode of remarkable gamma-ray flaring activity from FSRQ PKS 2023-07 during April 2016, as detected by both the AGILE (Astrorivelatore Gamma a Immagini LEggero) and Fermi satellites. An intensive multiwavelength campaign, triggered by Swift, covered the entire duration of the flaring activity, including the peak gamma-ray activity.}
{We report the results of multiwavelength observations of the blazar. We found that during the peak emission, the most energetic photon had an energy of 44 GeV, putting strong constraints on the opacity of the gamma-ray dissipation region. The overall spectral energy distribution (SED) is interpreted in terms of leptonic models for blazar jets, with the emission site located beyond the broad line region (BLR).}
{}

\keywords{}

\maketitle

\section{Introduction} \label{sec:intro}

Blazars, according to the unification scheme of \citet{urry1995}, are radio loud active galactic nuclei (AGNs) with jets oriented close to the line of sight. Their emission extends to the whole electromagnetic spectrum, from radio to very high-energy \grays. The variability timescales span from minutes (e.g., PKS 2155-304, \citealp{aharonian2007}) to years (e.g., BL Lacertae, \citealp{raiteri2013}). In general, the spectral energy distribution (SED) shows two main components related to non-thermal emission from relativistic jets: the first one due to synchrotron processes, the second  one - peaking at \gray energies - produced by inverse Compton (IC) scatterings. According to this leptonic scenario, the seed photons are assumed to come either from synchrotron emission itself (i.e., synchrotron self-Compton, SSC) for BL Lacertae objects, or from a source external to the jet (i.e., external Compton, EC) for flat-spectrum radio quasars (FSRQs). In this scenario, several potential sources can be responsible for the seed photon bath, for example, the direct disk emission, the reprocessed disk emission from the broad line regions (BLRs), or from the molecular torus. For the FSRQs, the $\gamma$-$\gamma$ absorption prevents the \gray emitting region from being inside the BLR cavity \citep{liu2006,bai2009}.

The quasar PKS~2023-07 (BZQJ2025-0735) is a FSRQ, located at redshift z=1.388, which was discovered as a bright radio source in the late 1960s \citep{pauliny1966,ekers1969}. In \grays, it was detected as a weak persistent source by both AGILE (Astrorivelatore Gamma a Immagini LEggero) (1AGL J2026-0732, \citealp{pittori2009}; 1AGLR J2027-0747, \citealp{verrecchia2013}) and \lat (Large Area Telescope) (3FGL J2025.6-0736, \citealp{acero2015}), but it can undergo flaring periods with strong flux variation (up to approximately one order of magnitude, see, e.g., \citealp{piano2016}).

\section{Data analysis} \label{data}

\subsection{AGILE data} \label{data:agile}

We carried out an analysis of the data collected by the \textit{GRID} (Gamma-Ray Imaging Detector, \citealp{barbiellini2002,prest2003}), the \gray silicon-tracker imager on board the AGILE satellite (for a detailed description of the AGILE payload: \citealp{tavani2009}).
The \grid is sensitive to \gray photons in the energy range 30 MeV -- 30 GeV. The point spread function (PSF) at 100 MeV and 400 MeV is $4.2^{\circ}$ and $1.2^{\circ}$ ($68\%$ containment radius), respectively \citep{sabatini2015}.
AGILE operated in a ``pointing'' mode for data-taking, characterized by fixed attitude observations, until November 2009 when the satellite entered a ``spinning'' mode, covering a large fraction of the sky with a controlled rotation of the pointing axis. In this current observing mode, typical two-day integration-time sensitivity (5$\sigma$) for sources outside the Galactic plane and photon energy above 100 MeV is ${\sim}2 \times 10^{-6}$ \flx.

We carried out the analysis of the \grid data above 100 MeV with the new \verb+Build_23+ scientific software, \verb+FM3.119+ calibrated filter and \verb+I0025+ response matrices. The consolidated archive, available from the ASI Data Center (\verb+ASDCSTDk+), was analyzed by applying South Atlantic Anomaly event cuts and $90^{\circ}$ Earth albedo filtering. Only incoming \gray events with an off-axis angle lower than $50^{\circ}$ were selected for the analysis. This is the most conservative standard configuration for the AGILE data analysis of incoming photons. Statistical significance and flux determination of the point sources were calculated by using the AGILE multi-source likelihood analysis (MSLA) software \citep{bulgarelli2012} based on the Test Statistic (TS) method as formulated by \citealp{mattox1996}. This statistical approach provides a detection significance assessment of a \gray source by comparing maximum-likelihood values of the null hypothesis (no source in the model) with the alternative hypothesis (point source in the field model).

\subsection{Fermi-LAT data} \label{data:fermi}

For the purpose of this work, we used the Science Tools provided by the \textit{Fermi} satellite team\footnote{{\tt http://fermi.gsfc.nasa.gov.}} on the PASS8 data around the position of PKS~2023-07. The version of the Science Tools used was \verb+v01r0p5+ with the \verb+P8R2_TRANSIENT010_V6+ instrument response function (IRF). The reader is referred to  \textit{Fermi} instrumental publications for further details about IRFs and other calibration details \citep{ackermann2012}.

For the maximum likelihood analysis, we adopted the current Galactic diffuse (\verb+gll_iem_v06.fits+) and isotropic emission models (\verb+iso_P8R2_SOURCE_V6_v06.txt+, \citealp{acero2016}), and the \lat four-year point source catalog (3FGL, \citealp{acero2015}). In the modeling of the data, the Galactic background and diffuse components remained fixed. We selected PASS8 FRONT and BACK transient class events with energies between 0.1 and 300 GeV. Among them, we limited the reconstructed zenith angle to be less than 105$^{\circ}$ to greatly reduce gamma rays coming from the limb of the Earth's atmosphere. We selected the good time intervals of the observations by excluding events that were taken while the instrument rocking angle was larger than 50$^{\circ}$. In the model for our source we used a power-law model for PKS~2023-07 and used the \verb+make3FGLxml.py+ tool to obtain a model for the sources within a 25$^{\circ}$ region of interest (ROI). To analyze the data we used the user contributed package \verb+Enrico+\footnote{{\tt https://github.com/gammapy/enrico/.}}. 

We divided each analysis into two steps: in the first one we leave all parameters of all the sources within a 10$^{\circ}$ ROI free, while the sources outside this ROI up to 25$^{\circ}$ have their parameters fixed. Then we run a likelihood analysis using the Minuit optimizer, which determines the spectral-fit parameters, and obtain a fit for all these sources. In the second step, we fix all the parameters of the sources in our model to the fitted values, except for our source of interest, and run again the likelihood analysis with the \verb+Newminuit+ optimizer to obtain a refined fit. At all times, the central target source PKS~2023-07 kept the spectral normalization free.

\subsection{Swift data} \label{data:swift}

The quasar PKS~2023-07 was observed by {\it Swift} \citep{gehrels2004} as a follow-up of the 
first AGILE flare starting from 2016 March 27 (ATel \#8879, \citealp{piano2016}), 
with both the X-ray Telescope \citep[XRT][see Table~\ref{xrtdatalog}]{burrows2005} 
and the UV/Optical Telescope \citep[UVOT][]{roming2005}. 
After the second AGILE flare on 2016 April 15 , the data were collected daily. 
The {\it Swift} data were uniformly processed and analyzed using the standard software 
(\verb+FTOOLS v6.20+), calibration (\verb+CALDB2 20170130+), and methods. 
The {\it Swift}/XRT data were processed and filtered with the task \verb+xrtpipeline+ (v0.13.3).  
Source events were extracted from a circular region with a radius of 20 pixels 
(one pixel corresponds to $2.36$\arcsec), background events 
from a circular region with a radius of 60 pixels in a source-free region nearby. 
A detection was performed by using {\em XIMAGE} in each observation and the measured count 
rates converted into fluxes by using the mean conversion factor derived from the spectral analysis 
of the combined observations 013-019.

The  {\it Swift}/UVOT data analysis was performed using the \verb+uvotimsum+ and \verb+uvotsource+ 
tasks within FTOOLS. The task \verb+uvotsource+ calculates the magnitude of the source through aperture 
photometry within a circular region centered on the source and applies the required corrections 
related to the specific detector characteristics. 
We adopted a circular region with a radius of 5\,\arcsec \, for the source and 15\,\arcsec\,  
for the background. In order to build the SED, we corrected the XRT and UVOT data for Galactic 
reddening according to the procedure described in \cite{raiteri2010}.

\subsection{Guillermo Haro Observatory data}

The near-infrared (NIR) photometry (J, H, Ks) was carried out with CANICA (CAnanea Near-Infrared CAmera) instrument at the 2.12 m telescope of the Guillermo Haro Astrophysical Observatory (OAGH) located in Cananea, Sonora, M{\'e}xico, whose technical details are
reported in \citet{carrasco2017}. Our standard reduction procedure includes proper flat
fielding and background subtraction from dithered images. Differential photometry is obtained
from field stars included in the 2MASS (2 Micron All Sky Survey) catalog. This approach allows us to obtain acceptable
data even on lightly clouded nights. The maximum error for the data reported here is 4\%,
although the formal error may be smaller.

\section{Results}\label{results}

We analyzed the \gray data above 100 MeV between 2016 March 15  UT 12:00:00 and 2016 April 30 UT 12:00:00. A 48-hour-bin light curve was calculated with a MSLA approach for both the \grid and \lat. For the AGILE light curve, we report $95\%$ confidence level (C. L.) flux upper limits (ULs) if TS $< 9$ (detection significance $\lesssim 3$) and flux values with the corresponding 1$\sigma$ statistical errors otherwise (TS $\geqslant 9$).  In the \lat analysis, we found that all the bins had a test statistic (TS) above 25 (meaning that all bins were detected with a confidence level above $\sim 5\sigma$). For both the \grid and \lat light curves, we assumed a power-law emission model with photon index 2.1, which represents the reference value in the AGILE analysis for unknown-spectrum sources (see, for example, \citealp{pittori2009}). We adopted the same photon index also in the Fermi-LAT analysis in order to preserve a uniform procedure.

\begin{figure*}[!htb]
   \begin{center}
    {\includegraphics[width=16cm]{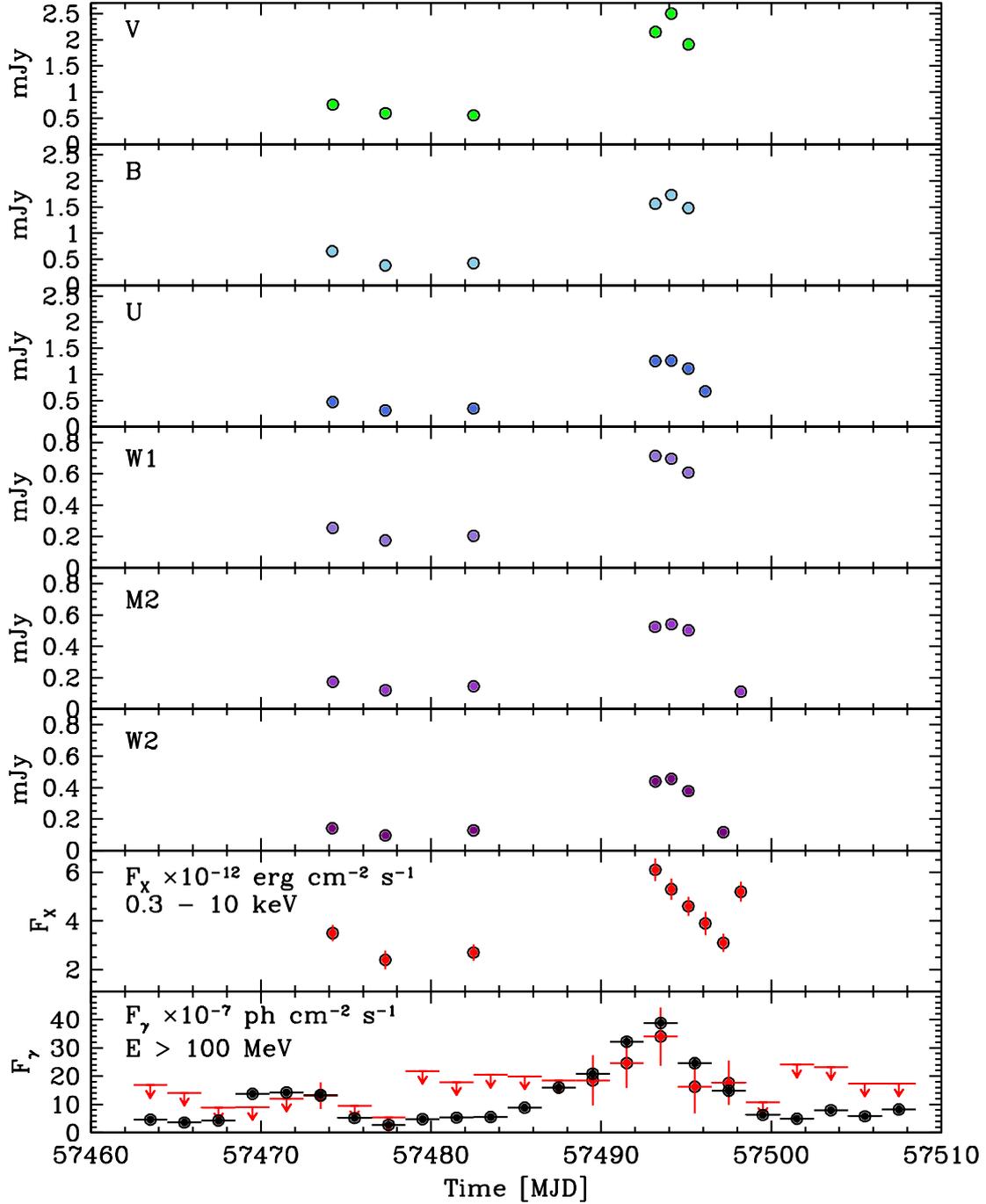}}
   \caption{From top to bottom: panels 1 to 6: UVOT data ($w1$, $m2$, $w2$, $u$, $b$, $v$ bands observed flux in units of mJy) acquired during the observations described in Table~\ref{xrtdatalog}; panel 7: {\it Swift}/XRT ($0.3-10.0$\,keV observed flux in units of $10^{-12}$\,erg\,cm$^{-2}$\,s$^{-1}$); last panel: \gray 48h-bin light curve ($> 100$\,MeV observed flux in units of $10^{-7}$\,photons\,cm$^{-2}$\,s$^{-1}$), AGILE data in red, \lat in black. The only NIR data from OAGH available during this time interval were taken on MJD $\sim$ 57498 (flux density = $(2.64 \pm 0.10)$ mJy in J band, $(4.16 \pm 0.19)$ mJy in H band, $(5.91 \pm 0.27)$ mJy in Ks band).}
    \label{pks2023:fig:mw_lc}
   \end{center}
\end{figure*}

As visible in Fig.~\ref{pks2023:fig:mw_lc}, the \gray emission is characterized by a first minor peak around March 27 (MJD $\sim$57474.5, see also ATel \#8879, \citealp{piano2016}). A second major peak in the \gray light curve is visible between April 10 UT 12:00:00 and April 20 UT 12:00:00 (MJD: 57488.5 -- 57498.5, see also ATel \#8932, \citealp{ciprini2016} and \#8960, \citealp{verrecchia2016}).
Remarkably, within four days centered around MJD 57493.5, \lat collected 13 gamma rays above $25\ \mathrm{GeV}/(1+z)$ and five gamma rays above $50\ \mathrm{GeV}/(1+z)$. This second major gamma-ray peak is in coincidence with a visible enhancement in the UV and X-ray emission. During this flare the UV/X-ray/\gray fluxes increase by a factor approximately three to four.

\begin{figure*}
  \begin{center}
    \includegraphics[width=12cm]{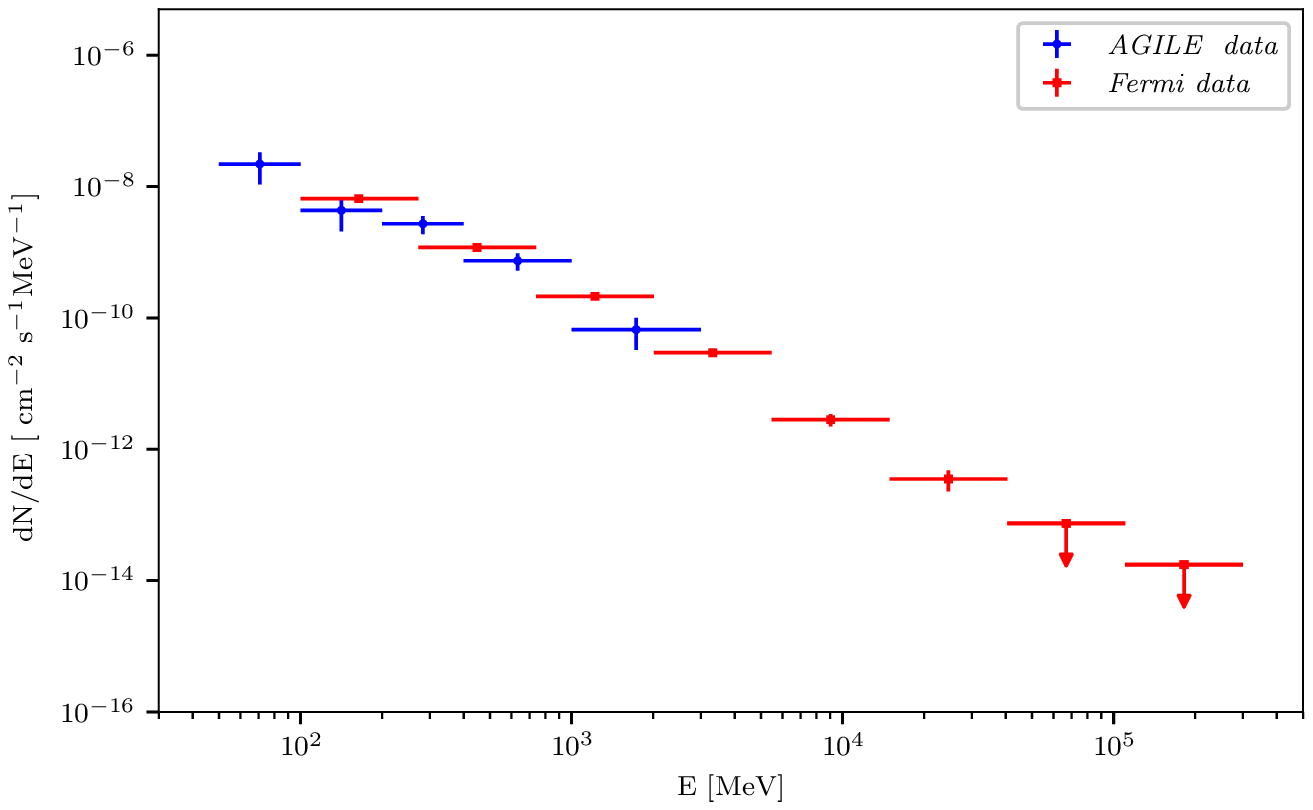}
    \caption{Differential spectrum for the second \gray flare, ten-day integration, from 2016 April 10 UT 12:00:00 to 2016 April 20 UT 12:00:00 (MJD: 57488.50 -- 57498.50). Blue points: \grid data, 50 MeV -- 3 GeV. Red points: \lat data, 100 MeV -- 300 GeV. }
    \label{fig:diff:spec}
  \end{center}
\end{figure*}

We obtained the differential spectrum of this second flare for both the \grid (between 50 MeV and 3 GeV) and \lat (between 100 MeV and 300 GeV). The results are depicted in Fig.~\ref{fig:diff:spec}. By fitting the spectra with a simple power law, we found that the inferred photon indices are fully compatible: $\gamma_{\, AGILE} = 1.8 \pm 0.1$, $\gamma_{\,Fermi} = 1.84 \pm 0.03$. During this flare \lat recorded a maximum photon energy of 44 GeV. We note that a power-law index value of $\sim$1.8 is much harder than the one reported in the 3FGL catalog: $\gamma_{\, 3FGL} = 2.18 \pm 0.03$.

Regarding the {\it Swift}/XRT data, a mean spectrum was extracted in the 0.3--10\,keV energy range from the combined observations collected during the second flare (ObsIDs 00036353013-019, for a total of 1075 counts, mean 0.3--10\,keV count rate $(8.8\pm0.3)\times10^{-2}$\,counts s$^{-1}$) and fitted with an absorbed power law. In the model, the hydrogen column density was fixed to $N_{\rm H}^{\rm Gal}= 3.24\times$10$^{20}$~cm$^{-2}$, corresponding to the Galactic value in the direction of the source \citep{kalberla2005}. 
We measured a power-law photon index $\Gamma=1.60\pm0.08$ ($\chi^{2}$/dof $=0.941/48$) and an observed 0.3--10\,keV flux of $(3.2\pm0.2)\times10^{-12}$ erg cm$^{-2}$ s$^{-1}$. 

In the NIR band (J, H, Ks bands) the source was observed by the OAGH at MJD $\sim$ 57498, 57553, 57580, 57688. We did not report the NIR data in the multiwavelength light curve in Fig.~\ref{pks2023:fig:mw_lc}, because only the first observation was performed during the descending phase of the flare (the other ones are outside the reported time interval). Nevertheless, the first observation represents the highest luminosity state (flux density = $(2.64 \pm 0.10)$ mJy in J band, $(4.16 \pm 0.19)$ mJy in H band, $(5.91 \pm 0.27)$ mJy in Ks band) ever detected from this blazar by OAGH. On the other hand, the observation on MJD $\sim$ 57688 shows the faintest NIR state of the source. These data are reported in the SED of Fig.~\ref{fig:sed_and_models} (see Section~\ref{discussion} for details).

 \begin{table*}
 \begin{center}
 \caption{Details of the \textit{Swift}/XRT observations of PKS 2023-07 between March and April 2016.}
 {\small
 \begin{tabular}{ccccc}
 \label{xrtdatalog}

  Sequence       &              Start time  (UT) &         End time   (UT)       & Exposure  & Flux \\
                       & (yyyy-mm-dd hh:mm:ss)  & (yyyy-mm-dd hh:mm:ss)  &        (s)    & ($10^{-12}$\,erg\,cm$^{-2}$\,s$^{-1}$) \\
       (1)           &                 (2)                 &            (3)                      &        (4)    & (5) \\
\hline\hline
00036353010        &       2016-03-27 03:25:04     &       2016-03-27 06:53:56     &       2984   & 3.50 $\pm$ 0.34\\
00036353011        &       2016-03-30 06:35:08     &       2016-03-30 08:30:54     &       1979   & 2.40 $\pm$ 0.38\\
00036353012        &       2016-04-04 09:28:11     &       2016-04-04 14:24:55     &       2956   & 2.70 $\pm$ 0.34\\
00036353013        &       2016-04-15 03:32:56     &       2016-04-15 05:24:05     &       2665   & 6.10 $\pm$ 0.46\\
00036353014        &       2016-04-16 02:06:45     &       2016-04-16 03:54:55     &       2924   & 5.30 $\pm$ 0.44\\
00036353015        &       2016-04-17 00:33:34     &       2016-04-17 05:33:54     &       2876   & 4.60 $\pm$ 0.39\\
00036353016        &       2016-04-18 01:56:53     &       2016-04-18 03:42:42     &       2610   & 3.90 $\pm$ 0.46\\
00036353018        &       2016-04-19 03:12:53     &       2016-04-19 05:16:53     &       2402   & 3.10 $\pm$ 0.38\\
00036353019        &       2016-04-20 03:27:34     &       2016-04-20 06:22:55     &       3064   & 5.20 $\pm$ 0.41\\ 
   \end{tabular}
   }
   \end{center}
   \end{table*}

\section{Discussion} \label{discussion}

The multiwavelength data available for PKS 2023-07 allowed us to build-up and interpret its SED. We modeled the spectra related to the April 2016 flare: AGILE and \lat data from 2016 April 10 UT 12:00:00 to 2016 April 20 UT 12:00:00 (MJD: 57488.50 -- 57498.50),  and \textit{Swift}-UVOT and \textit{Swift}-XRT data from 2016 April 15 UT 03:32:56 and 2016 April 20 UT 06:22:55 (MJD: 57493.15 -- 57498.27).
Remarkably, the gamma-ray data show a hard $\rm \gamma$-ray spectrum extending up to $\sim 50 \rm GeV$ (with uncommon emission above $50\ \mathrm{GeV}/(1+z)$ within four days centered around MJD 57493.5, as reported in Section \ref{results}), providing important information about the gamma-ray absorption in the high-energy emitting region. Several gamma-ray flaring states of FSRQs show a high-energy emission above $25\ \mathrm{GeV}/(1+z)$ \citep[see., e.g., ][]{aleksic2011,tavecchio2011,abdo2010,hayashida2012,ghisellini2013,nalewajko2012,pacciani2012,tavecchio2013,pacciani2014,ahnen2015,abeysekara2015}.

\begin{figure*}
\begin{center}

 \subfigure[SED and model \emph{a}]
   {\includegraphics[width=15cm]{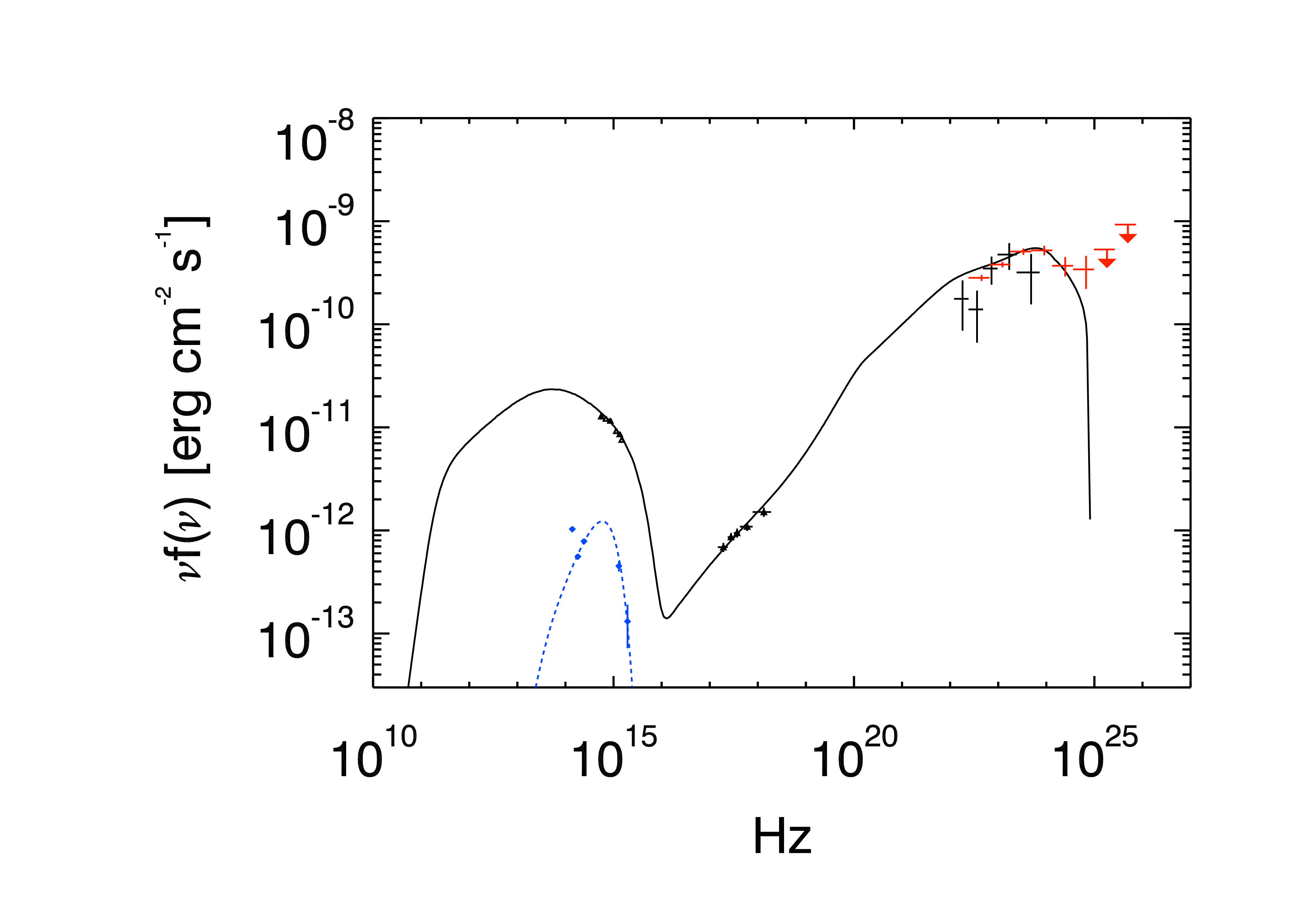}}
 \hspace{1mm}
 \subfigure[SED and model \emph{b}]
   {\includegraphics[width=15cm]{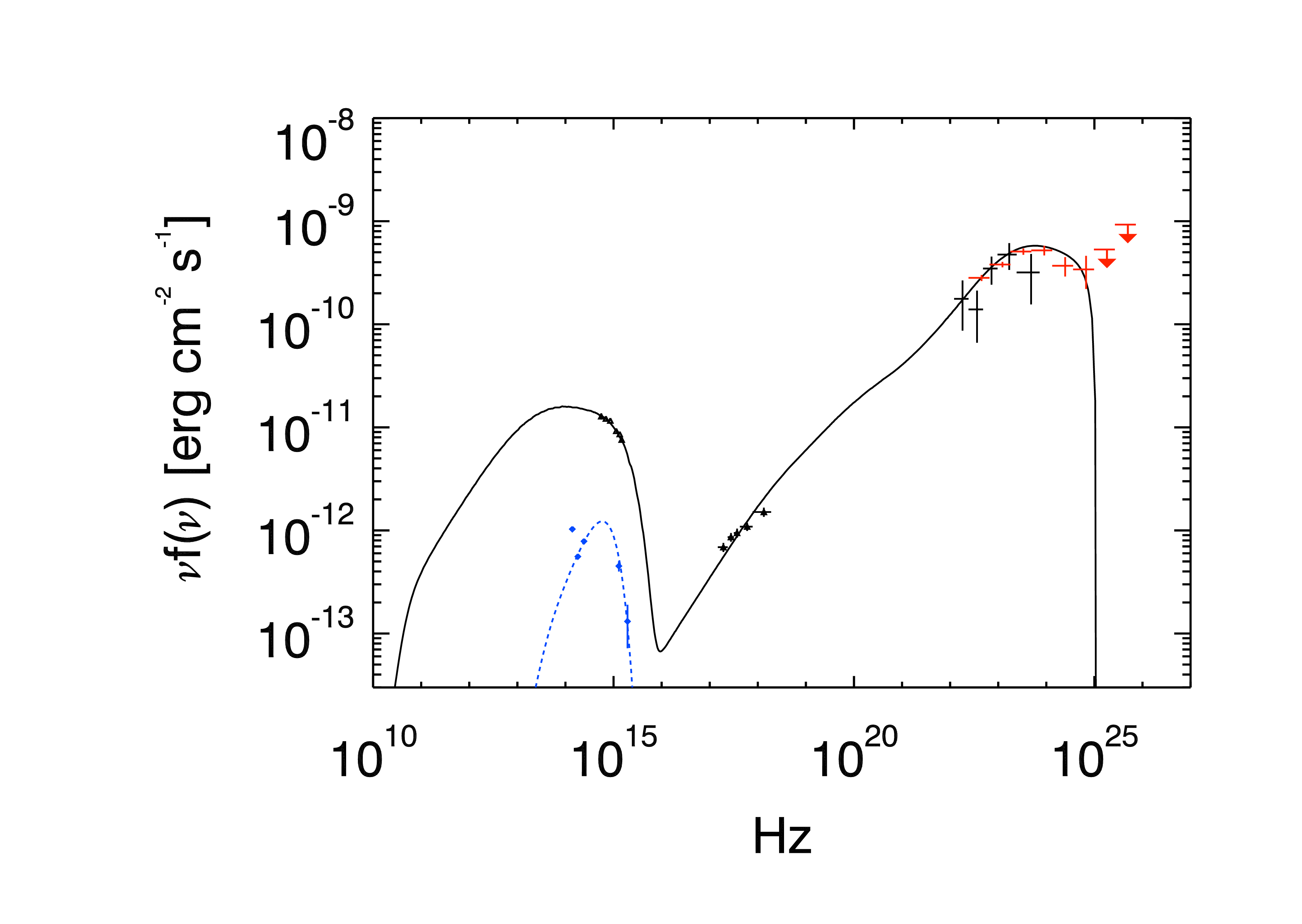}}
 \caption{Spectral energy distribution for the brightest flare in April 2016 of PKS 2023-07. AGILE (black points) and \lat (red point) data from 2016 April 10 UT 12:00:00 to 2016 April 20 UT 12:00:00 (MJD: 57488.50 -- 57498.50), and \textit{Swift}-UVOT and \textit{Swift}-XRT data from 2016 April 15 UT 03:32:56 and 2016 April 20 UT 06:22:55 (MJD: 57493.15 -- 57498.27). The low activity period observed in NIR on MJD $\sim$ 57688 is reported in gray together with archival GALEX data, and with the Shakura Sunyaev accretion disk model. Top (bottom) panel: SED model \emph{a} (SED model \emph{b}) for the flaring period is shown.}
\label{fig:sed_and_models}

\end{center}
 \end{figure*}

The opacity argument from the broad line region (BLR) \citep{liu2006,boettcher2016} reveals
that the gamma-ray emission site cannot be located closer than the internal
radius of the BLR (assuming a spherical shell geometry for the BLR). Therefore, it provides important information about the site where gamma-rays are emitted. In order to obtain an estimate of the gamma-ray opacity along the jet, first the disk luminosity has to be derived. 
Actually, optical spectroscopic data are not available for this source,
preventing us from obtaining the accretion disk luminosity as proposed in \cite{celotti1997} and then constrain the gamma-ray opacity as a function of the distance from the supermassive black hole (SMBH).
On the other hand, we found that NIR observations performed on MJD $\sim$ 57688, after the flaring period, at the OAGH revealed the source at its lowest activity period ever detected in NIR.
These NIR data show an upturn of source flux in the $J$ band with respect to that in the $K_S$ and $H$ filters (see blue circles in Fig.~\ref{fig:sed_and_models}), likely linked to the accretion disk emission.
Combining these  data with those in NUV and FUV bands taken from the GALEX (Galaxy Evolution Explorer) archive, we  estimated a disk
luminosity of  $ 1-2 \times 10^{46}$ erg/s assuming a standard Shakura Sunyaev
accretion disk model. 
The data and disk model are shown in Fig.~\ref{fig:sed_and_models} together with
the broad band SED for the flaring period.\\

Adopting the \cite{liu2006} BLR modeling, and a Shakura-Sunyaev accretion disk with a luminosity of $L_{disk}\ =\ 1.5\cdot10^{46}$ erg/s, we expect the emission lines of the BLR to be opaque to gamma-ray radiation above $25\ \mathrm{GeV}/(1+z)$.
In particular, the opacity to 50 GeV gamma rays (host galaxy frame) is $\tau \simeq 3.6$ for an emitting region located at the internal BLR radius.
 
We adopted the following geometry of the SMBH environment as proposed for the SED modeling of FSRQs in \cite{sikora2009} \cite{ghisellini2009}: a BLR with an internal radius $R_{BLR}\ = 3.9 \cdot 10^{17}$ cm, according to 
R$_{BLR} \propto L_{disk}^{0.5}$ scaling relation \citep{bentz2006,kaspi2007} and a dusty torus with a radius of $10^{19}$ cm re-emitting $\sim$30\% of the incoming accretion disk radiation.
We constrained the size  of the emitting region
 $R_{blob}\ \sim\ 10^{17}$ cm to match the variability timescale of approximately four days shown in  the light
curves of Fig.~\ref{pks2023:fig:mw_lc}. Then, we modeled the SED in the framework of leptonic models.
We describe the lepton population with a distribution
$\propto \frac{(\gamma/\gamma_b)^{-s_1}}{1+(\gamma/\gamma_b)^{-s_1+s_2}}$,
with $\gamma$ the electron Lorentz factor ($\gamma_{min} < \gamma < \gamma_{max}$), and $\gamma_b$ the break Lorentz factor.\\
Light curves in Fig.~\ref{pks2023:fig:mw_lc} suggest simultaneous emission in optical and
\grays. In such a case,
we can use the simultaneous spectrum of both synchrotron and external Compton emissions
to investigate the Klein-Nishina (KN) regime.
A sizable KN suppression is expected if the broad line photons interact
face-on with leptons of the moving plasmoid.\\ 

In our leptonic framework, we found two possible solutions for the emission site.
A possible location is at the edge of the BLR, where both the BLR and torus
radiation fields effectively contribute to the external Compton process.
With this geometry, BLR photons come from behind the jet, and Klein-Nishina suppression
is slightly reduced in the high-energy range.
The inferred magnetic field is $\sim$ 0.46 G. This scenario is reported as model \textit{a} in Fig.~\ref{fig:sed_and_models} and the corresponding parameters are on the left side of Table \ref{tab:models}.
\\
Alternatively, the dissipation region is  located several parsecs away from
the SMBH, with the Torus dominating among the external radiation
fields for the external Compton in gamma rays, and a weak magnetic field
of 0.043 G responsible for the synchrotron emission.
This second solution  describes well the high-energy emission. This scenario is reported as model \textit{b} in Fig.~\ref{fig:sed_and_models}, the corresponding parameters are on the right side of Table \ref{tab:models}.
The lack of NIR data simultaneous with the flare, in the spectral region where the synchrotron component peaks, prevented us from drawing a firm conclusion on the gamma-ray site because of the uncertainty on the value of magnetic field. On the other hand, the modeling of the multiwavelength SED seems to exclude a gamma-ray emitting region in the close vicinity of the SMBH, that is, within the BLR. Assuming that the adopted modeling correctly describes the BLR and dusty torus radiation field at any distance from the SMBH, the solutions found for $R_{diss}$ for both models \textit{a} and \textit{b} are poorly sensitive to change of the other model parameters.  In fact, for both models, $R_{diss}$ is found in the steep descending region of the energy density profile of the corresponding external radiation field \citep[see, e.g.,][]{ghisellini2009}, implying that the modeled \gray emission is very sensitive to $R_{diss}$ variation. Therefore, given the very narrow range allowed for $R_{diss}$, we exclude intermediate dissipation regions far enough away from the current values assumed for models \textit{a} and \textit{b}.

\begin{table}
\caption{Model parameter for the spectral energy density in the two assumptions of a blob dissipating just beyond the BLR (R$_{diss} = 0.2$ pc, model \textit{a}) or far away from the SMBH (R$_{diss} = 6.8$ pc, model \textit{b}). The parameter f$_{BLR}$ is the fraction of the disk luminosity that is reprocessed by the BLR and re-emitted as broad emission lines, f$_{Torus}$ is the fraction of the disk luminosity that is reprocessed by the dusty torus, $\epsilon_{accr}$ is the accretion efficiency, and $R_S$ is the Schwarzschild radius of the SMBH.}
\label{tab:models}
\centering
\begin{tabular}{|l|c|c|}
\hline
model                          & \textit{\textbf{a}}          & \textit{\textbf{b}}         \\ \hline   \hline                  
R$_{diss}$ (pc)            & 0.29                              & 6.8                                \\ \hline  
Blob radius   (cm)         & 7.7$\times 10^{16}$     & 1.4$\times 10^{17}$     \\ \hline
L$_{d}$ (erg/s)           & \multicolumn{2}{c|}{1.5$\times 10^{46}$}              \\ \hline      
R$_{BLR}$ (pc)           & \multicolumn{2}{c|}{0.13}                                        \\ \hline
R$_{Torus}$ (pc)        & \multicolumn{2}{c|}{3.24}                                        \\ \hline
f$_{BLR}$                    & \multicolumn{2}{c|}{0.1}                                         \\ \hline
f$_{Torus}$                  & \multicolumn{2}{c|}{0.3}                                        \\ \hline
$\epsilon_{accr}^{0.5} \cdot R_S$  (cm)     & \multicolumn{2}{c|}{$6\cdot 10^{14}$}\\            \hline 
$\Gamma_{bulk}$                                         & 20                             & 30                            \\ \hline
angle of view (deg)                                        & 2                               & 2                              \\ \hline
$\gamma_{min}$                                          & 1                               & 1                              \\ \hline
$\gamma_{max}$                                         & 6.1$\times 10^3$      & 1.5$\times 10^4$    \\ \hline
$\gamma_{break}$                                       & 0.95$\times 10^3$    & 2.00$\times 10^3$  \\ \hline
density at $\gamma_{break}$  (cm$^{-3}$)  & 3.9$\times 10^{-2}$  & 2.0$\times 10^{-3}$ \\ \hline                               
$s_1$                                                          & 1.1                            & 1.5                            \\ \hline
$s_2$                                                          & 3.5                            & 3.5                            \\ \hline
$B$     (G)                                                    & 4.6$\times 10^{-1}$  & 4.3$\times 10^{-2}$ \\ \hline  \hline

\end{tabular}

\end{table}

\begin{acknowledgements}
The authors are grateful to the anonymous referee for her or his stimulating comments on the manuscript.
Research partially supported by the ASI grant no. I/028/12/5.
L. Pacciani acknowledges the contribution from the grant INAF CTA–SKA.

\end{acknowledgements}


\begin{thebibliography}{99}


\bibitem[Abdo et al.(2010)]{abdo2010}
Abdo , A. A. et al., 2010, Nature, 463, 919;

\bibitem[Abeysekara et al.(2015)]{abeysekara2015}
Abeysekara, A. U., et al, 2015, ApJL, 815, 22    

\bibitem[Acero et al.(2015)]{acero2015}
Acero, F., Ackermann, M., Ajello, M., et al.\ 2015, \apjs, 218, 23 

\bibitem[Acero et al.(2016)]{acero2016}
Acero, F., Ackermann, M., Ajello, M., et al.\ 2016, \apjs, 223, 26 

\bibitem[Ackermann et al.(2012)]{ackermann2012}
Ackermann, M., Ajello, M., Albert, A., et al.\ 2012, \apjs, 203, 4 

\bibitem[Aharonian et al.(2007)]{aharonian2007}
Aharonian, F., Akhperjanian, A.~G., Bazer-Bachi, A.~R., et al.\ 2007, \apjl, 664, L71 

\bibitem[Ahnen et al.(2015)]{ahnen2015}
Ahnen, M. L., et al., 2015, ApJL, 815, 23 

\bibitem[Aleksic et al.(2011)]{aleksic2011}
Aleksic, J. et al., 2011, ApJL, 730, 8

\bibitem[Bai et al.(2009)]{bai2009}
Bai, J.~M., Liu, H.~T., \& Ma, L.\ 2009, \apj, 699, 2002 

\bibitem[Barbiellini et al.(2002)]{barbiellini2002}
Barbiellini, G., Fedel, G., Liello, F., et al.\ 2002, Nuclear Instruments and Methods in Physics Research A, 490, 146

\bibitem[Bentz et al.(2006)]{bentz2006}
Bentz, M. C., Peterson B. M., Pogge R. W., Vestergaard M., Onken C., 2006, ApJ, 6, 133

\bibitem[B{\"o}ttcher \& Els(2016)]{boettcher2016}
B{\"o}ttcher, M. \& Els, P. 2016, ApJ, 821, 102

\bibitem[Bulgarelli et al.(2012)]{bulgarelli2012}
Bulgarelli, A., Chen, A.~W., Tavani, M., et al.\ 2012, \aap, 540, A79

\bibitem[Burrows et al.(2005)]{burrows2005}
Burrows, D.~N., Hill, J.~E., Nousek, J.~A., et al.\ 2005, \ssr, 120, 165 

\bibitem[Carrasco et al.(2017)]{carrasco2017}
Carrasco, L., Hern{\'a}ndez Utrera, O., V{\'a}zquez, S., et al.\ 2017, \rmxaa, 53, 497

\bibitem[Celotti et al.(1997)]{celotti1997}
Celotti, A., Padovani, P., \& Ghisellini, G. 1997, MNRAS, 286, 415

\bibitem[Ciprini \& Fermi Large Area Telescope Collaboration(2016)]{ciprini2016}
Ciprini, S., \& Fermi Large Area Telescope Collaboration 2016, The Astronomer's Telegram, 8932

\bibitem[Ekers(1969)]{ekers1969}
Ekers, J.~A.\ 1969, Australian Journal of Physics Astrophysical Supplement, 7,  

\bibitem[Gehrels et al.(2004)]{gehrels2004}
Gehrels, N., Chincarini, G., Giommi, P., et al.\ 2004, \apj, 611, 1005 

\bibitem[Ghisellini \& Tavecchio(2009)]{ghisellini2009}
Ghisellini, G. and Tavecchio, F., 2009, MNRAS, 397, 985

\bibitem[Ghisellini et al.(2013)]{ghisellini2013}
Ghisellini, G. et al., 2013, MNRASL, 432, 66

\bibitem[Hayashida et al.(2012)]{hayashida2012}
Hayashida, M. et al., 2012, ApJ, 754, 114

\bibitem[Kalberla et al.(2005)]{kalberla2005}
Kalberla, P.~M.~W., Burton, W.~B., Hartmann, D., et al.\ 2005, \aap, 440, 775 

\bibitem[Kaspi et al.(2007)]{kaspi2007}
Kaspi S. et al., ApJ, 659, 997

\bibitem[Liu \& Bai(2006)]{liu2006}
Liu, H. T., \& Bai, J. M. 2006, ApJL, 653, L1089

\bibitem[Mattox et al.(1996)]{mattox1996}
Mattox, J.~R., Bertsch, D.~L., Chiang, J., et al.\ 1996, \apj, 461, 396 

\bibitem[Nalewajko et al.(2012)]{nalewajko2012}
Nalewajko K. et al., 2012, ApJ, 760, 69

\bibitem[Pacciani et al.(2012)]{pacciani2012}
Pacciani L. et al., 2012, MNRAS, 425, 2015

\bibitem[Pacciani et al.(2014)]{pacciani2014}
Pacciani L. et al., 2014, ApJ, 790, 45

\bibitem[Pauliny-Toth et al.(1966)]{pauliny1966}
Pauliny-Toth, I.~I.~K., Wade, C.~M., \& Heeschen, D.~S.\ 1966, \apjs, 13, 65 

\bibitem[Piano et al.(2016)]{piano2016}
Piano, G., Bulgarelli, A., Tavani, M., et al.\ 2016, The Astronomer's Telegram, 8879

\bibitem[Pittori et al.(2009)]{pittori2009}
Pittori, C., Verrecchia, F., Chen, A.~W., et al.\ 2009, \aap, 506, 1563 

\bibitem[Prest et al.(2003)]{prest2003}
Prest, M., Barbiellini, G., Bordignon, G., et al.\ 2003, Nuclear Instruments and Methods in Physics Research A, 501, 280 

\bibitem[Raiteri et al.(2010)]{raiteri2010}
Raiteri, C.~M., Villata, M., Bruschini, L., et al.\ 2010, \aap, 524, A43 

\bibitem[Raiteri et al.(2013)]{raiteri2013}
Raiteri, C.~M., Villata, M., D'Ammando, F., et al.\ 2013, \mnras, 436, 1530 

\bibitem[Roming et al.(2005)]{roming2005}
Roming, P.~W.~A., Kennedy, T.~E., Mason, K.~O., et al.\ 2005, \ssr, 120, 95 

\bibitem[Sabatini et al.(2015)]{sabatini2015}
Sabatini, S., Donnarumma, I., Tavani, M., et al.\ 2015, \apj, 809, 60 

\bibitem[Sikora et al.(2009)]{sikora2009}
Sikora, M. et al., 2009, ApJ, 704, 38

\bibitem[Tavani et al.(2009)]{tavani2009}
Tavani, M., Barbiellini, G., Argan, A., et al.\ 2009, \aap, 502, 995 

\bibitem[Tavecchio et al.(2011)]{tavecchio2011}
Tavecchio, F. et al., 2011, A\&A, 534, 86;

\bibitem[Tavecchio et al.(2013)]{tavecchio2013}
Tavecchio, F., Pacciani, L., Donnarumma, I., et al.\ 2013, \mnras, 435, L24 

\bibitem[Urry \& Padovani(1995)]{urry1995}
Urry, C.~M., \& Padovani, P.\ 1995, \pasp, 107, 803 

\bibitem[Verrecchia et al.(2013)]{verrecchia2013}
Verrecchia, F., Pittori, C., Chen, A.~W., et al.\ 2013, \aap, 558, A137 

\bibitem[Verrecchia et al.(2016)]{verrecchia2016}
Verrecchia, F., Tavani, M., Lucarelli, F., et al.\ 2016, The Astronomer's Telegram, 8960


\end{thebibliography}
\end{document}